\documentclass{physeauth}
\usepackage{amssymb}
\usepackage{amsmath}
\usepackage{graphicx}

\begin{document}

\begin{frontmatter}

\title{Length scale dependent diffusion in the Anderson model at high temperatures}

\author[uos]{Robin Steinigeweg}, 
\author[uos]{Jochen Gemmer \thanksref{email2}}

\address[uos]{Fachbereich Physik, Universit\"at Osnabr\"uck,
              Barbarastrasse 7, D-49069 Osnabr\"uck, Germany}

\thanks[email2]{Corresponding author. E-mail: jgemmer@uos.de}

\begin{abstract}
We investigate a single particle on a $3$-dimensional, cubic
lattice with a random on-site potential ($3$D Anderson model). We
concretely address the question whether or not the dynamics of the
particle is in full accord with the diffusion equation. Our
approach is based on the time-convolutionless (TCL) projection
operator technique and allows for a detailed investigation of this
question at high temperatures. It turns out that diffusive
dynamics is to be expected for a rather short range of
wavelengths, even if the amount of disorder is tuned to maximize
this range. Our results are partially counterchecked by the
numerical solution of the full time-dependent Schr\"odinger
equation.
\end{abstract}


\end{frontmatter}

Since it had been suggested by P.~W.~Anderson, the Anderson model
served as a paradigm for transport in disordered systems
\cite{anderson1958,abouchacra1973,lee1985,kramer1993,erdos2007,schwartz2007}.
In its probably simplest form the Hamiltonian may be written as
\begin{equation}
H = \sum_{\bf r} \epsilon^{}_{\bf r} \, a^\dagger_{\bf r} \,
a^{}_{\bf r} + \sum_{\text{NN}} a^\dagger_{\bf r} \, a^{}_{\bf r'}
\; , \label{anderson}
\end{equation}
where $a_{\bf r}$, $a^\dagger_{\bf r}$ are the usual annihilation,
respectively creation operators; $\bf r$ labels the sites of a
$d$-dimensional lattice; and NN indicates a sum over nearest
neighbors. The $\epsilon_{\bf r}$ are independent random numbers,
e.g., Gaussian distributed numbers with mean $\langle
\epsilon_{\bf r} \rangle = 0$ and variance $\langle \epsilon_{\bf
r} \, \epsilon_{\bf r'} \rangle = \delta_{{\bf r}, {\bf r}'} \,
\sigma^2$. Thus, the first sum in Eq.~(\ref{anderson}) describes a
random on-site potential and hence disorder.

The phenomenon of localization, including localization lengths,
has intensively been studied in this system
\cite{anderson1958,abouchacra1973,lee1985,kramer1993}. For the
lower dimensional cases $d = 1$ and $d = 2$ (in the thermodynamic
limit, i.e., with respect to the infinite length scale) an
insulator results for arbitrary (non-zero) values of $\sigma$,
see, e.g., Ref.~\cite{kramer1993}. Of particular interest is the
$3$-dimensional case. Here, at zero temperature, increasing
disorder induces a metal-to-insulator transition at the infinite
length scale \cite{lee1985,kramer1993}. However, with respect to
finite length scales the following transport types are generally
expected: i.)~ballistic on a scale below some, say,  mean free
path; ii.)~possibly diffusive on a scale above this mean free path
but below the localization length; iii.)~localized (isolating) on
a scale above the localization length. In the above transition
decreasing disorder is viewed to reduce the size of the third
regime, until it vanishes.

Here, other than most of the pertinent literature, we do not focus
on the mere existence of a finite localization length. We rather
concentrate on the size of the intermediate regime and the
dynamics within. We demonstrate that it is indeed diffusive
(rather than subdiffusive, superdiffusive, or anything else). In
principle, for long localization lengths (or no localization) this
regime could be very large. But the results presented in the paper
at hand indicate that it is not, at least not in the limit of high
temperatures. Investigations in that direction (but not for $d=3$)
are also performed in Refs.~\cite{schwartz2007,lherbier2008}.

Our approach is based on the time-convolutionless (TCL) projection
operator technique \cite{breuer2007} which has already been
applied to the transport properties of similar models without
disorder
\cite{steinigeweg2007-1,michel2007-1,steinigeweg2006-2,michel2006}.
In its simplest form (which we apply here) this method is
restricted to the infinite temperature limit. This implies that
energy dependences are not resolved, i.e., our results are to be
interpreted as results on an overall behavior of all energy
regimes.

\begin{figure}[htb]
\centering
\includegraphics[width=6.5cm]{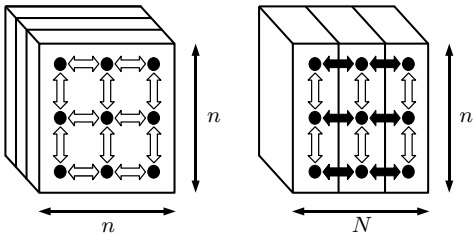}
\caption{A $3$-dimensional lattice which consists of $N$ layers
with $n \times n$ sites each. Only next-neighbor hoppings are
taken into account. Intra-layer hoppings are specified by a
constant $\alpha = 1$ (white arrows), inter-layer hoppings by
another constant $\beta$ (black arrows). } \label{model}
\end{figure}

As shown in Fig.~\ref{model}, we consider a $3$-dimensional
lattice consisting of $N$ layers with $n \times n$ sites each. The
Hamiltonian of our model is almost identical to
Eq.~(\ref{anderson}) with one single exception: All hopping terms
that correspond to hoppings between layers (black arrows in
Fig.~\ref{model}) are multiplied by some constant $\beta$. This is
basically done due to technical reasons, see below. However, for
$\beta=1$ the Hamiltonian reduces to the standard Anderson
Hamiltonian (\ref{anderson}).

We now establish a ``coarse-grained'' description in terms of
subunits: At first we take all those terms of the Hamiltonian
which only contain the sites of the $\mu$th layer in order to form
the local Hamiltonian $h_\mu$ of the subunit $\mu$. Thereafter all
those terms which contain the sites of adjacent layers $\mu$ and
$\mu+1$ are taken in order to form the interaction $\beta \,
v_\mu$ between neighboring subunits $\mu$ and $\mu+1$. Then the
total Hamiltonian may be also written as $H = H_0 + \beta \, V$,
\begin{equation}
H_0 = \sum_{\mu = 0}^{N-1} h_\mu \, , \quad V = \sum_{\mu =
0}^{N-1} v_\mu \; , \label{hamiltonian}
\end{equation}
where we employ periodic boundary conditions, e.g., we identify
$\mu = N$ with $\mu = 0$. The above introduction of the additional
parameter $\beta$ thus allows for the independent adjustment of
the ``interaction strength''. We are going to work in the
interaction picture. The hence indispensable eigenbasis of $H_0$
may be found from the diagonalization of disconnected layers.

By $\Pi_\mu$ we denote the particle number operator of the $\mu$th
subunit, i.e.,  the sum of $a^\dagger_{\bf r} \, a^{}_{\bf r}$
over all ${\bf r}$ of the $\mu$th layer. Since $[\sum_\mu
\Pi_\mu,H] = 0$, the one-particle subspace may be analyzed
separately, which will be done throughout this work.

The total state of the system is naturally represented by a
time-dependent density matrix $\rho(t)$. Consequently, the
quantity $P_\mu(t) \equiv \text{Tr} \{ \, \rho(t) \, \Pi_\mu \,
\}$ is the probability for locating the particle somewhere within
the $\mu$th subunit. The consideration of these ``coarse-grained''
probabilities corresponds to the investigation of transport along
the direction which is perpendicular to the layers. Instead of
simply characterizing whether or not there is transport at all, we
analyze the full dynamics of the $P_\mu$.

Those dynamics may be called diffusive, if the $P_\mu$ fulfill a
discrete diffusion equation
\begin{equation}
\dot{P}_\mu = \kappa \, ( \, P_{\mu+1} + P_{\mu-1} - 2 \, P_\mu \, ) \;
\label{diffusion1}
\end{equation}
with some diffusion constant $\kappa$. A decoupled form of this
equation is routinely derived by a transformation onto, e.g.,
cosine-shaped Fourier modes, that is, Eq.~(\ref{diffusion1})
yields
\begin{equation}
\dot{F}_q = -2 \, (1 - \cos q) \, \kappa \, F_q \; , \quad F_q
\equiv C_q \sum \label{diffusion2}_{\mu = 0}^{N - 1} \cos (q \,
\mu) \, P_\mu
\end{equation}
with $q = 2 \pi \, k / \, N$, $k = 0, 1, \ldots, N / \, 2$ and
$C_q$ being a yet arbitrary constant. Thus, a system is said to
behave diffusively at some wave number $q$, i.e., on some length
scale $l \equiv 2 \pi / q$, if the corresponding modes $F_q$ relax
exponentially.

For our purposes, the comparison of the resulting quantum dynamics
with Eq.~(\ref{diffusion2}), it is convenient to express the modes
$F_q$ as expectation values of mode operators $\Phi_q$,
\begin{equation}
F_q(t) = \text{Tr} \{ \, \rho(t) \, \Phi_q \, \} \; , \quad \Phi_q \equiv C_q
\sum_{\mu = 0}^{N - 1} \cos (q \, \mu) \, \Pi_\mu \; , \label{mode}
\end{equation}
where the $C_q$ are now chosen such that $\text{Tr} \{ \, \Phi_q^2 \, \} = 1$.

Our approach for the analysis of the $F_q$ is based on the TCL
projection operator technique, see Ref.~\cite{breuer2007}. For the
application of this technique we have to define a suitable
projection superoperator $\bf P$. Here, we choose
\begin{equation}
{\bf P} \, \rho(t) \equiv \text{Tr}\{ \, \rho(t) \, \Phi_q \, \}
\, \Phi_q = F_q(t) \, \Phi_q \; . \label{projector}
\end{equation}
For initial states $\rho(0)$ which satisfy ${\bf P} \, \rho(0) =
\rho(0)$ [which essentially means that we consider the decay of
harmonic density waves] the method eventually leads to a
differential equation of the form
\begin{equation}
\label{alltcl} \dot{F}_q(t) = \underbrace{(\beta^2 \,
\Gamma_{2,q}(t) + \beta^4 \, \Gamma_{4,q}(t) +
\ldots)}_{\Gamma_q(t)} \, F_q(t) \, .
\end{equation}
Note that $\rho(0)$ is not restricted to any energy subspaces and
accordingly corresponds to a state of high temperature.
Apparently, the dynamics of $F_q$ is controlled by a
time-dependent decay rate $\Gamma_q(t)$. This rate is given in
terms of a systematic perturbation expansion in powers of the
inter-layer coupling. (Concretely calculating the $\Gamma_{i,q}$
reveals that all odd orders vanish for this model.) At first we
concentrate on the analysis of Eq.~(\ref{alltcl}) to lowest
(second) order, however, below considering the fourth order will
account for localization. The TCL formalism yields
$\Gamma_{2,q}(t)= \int_0^t \text{d} \tau \, f_q(\tau)$, where
$f_q(\tau)$ denotes the two-point correlation function
\begin{equation}
f_q(\tau) = \beta^2 \; \text{Tr} \Big \{ [ \, V(t), F_q \, ] \,
[ \, V(t'), F_q \, ] \Big \} \, , \; \tau \equiv t-t' \; .
\label{correlation1}
\end{equation}
Here and in the following the time-dependencies of operators are to be
understood w.r.t. to the Dirac picture. A rather lengthy but
straightforward analysis shows that
Eq.~(\ref{correlation1}) significantly simplifies under the
following assumption: The autocorrelation functions $\text{Tr} \{ \, v_\mu(t) \, v_\mu(t') \, \}$
of the local interactions $v_\mu$ should
only depend negligibly on the layer number $\mu$ (during some
relevant time scale). Simple numerics indicate that this
assumption is well fulfilled (for the choices of $\sigma$
discussed here), once the layer sizes exceed ca.~$30 \times 30$.
Hence, first investigations may be based on the consideration of
an arbitrarily chosen junction of two layers, the interaction in
between we label by $\mu = 0$, i.e., we may consider $v_0$ in the
following. Exploiting this assumption reduces Eq.~(\ref{correlation1})
to
\begin{eqnarray}
& f_q(\tau) \approx -W f(\tau) \, , \\
& W \equiv 2 \, (1 - \cos q) \, \beta^2 \, , \; f(\tau) \equiv
1/n^2 \, \text{Tr} \{ \, v_0(t) \, v_0(t') \, \} \, . \nonumber
\label{correlation2}
\end{eqnarray}
(Note that the above approximation is exact for identical
subunits, see Ref.~\cite{steinigeweg2007-1}.)

Direct numerical computation shows that $f(\tau)$ looks like a
standard correlation function, i.e., it decays completely before
some time $\tau_C$. Of primary interest surely is
$\tilde{\Gamma}_2(t)  \equiv \int_0^t \text{d} \tau \, f(\tau)$.
Numerics indicate that neither $\tau_C$ nor $\gamma \equiv
\tilde{\Gamma}_2(t)$, $t > \tau_C$ [the area under the initial
peak of $f(\tau)$] depend substantially on $n$ (again for $n >
30$). Thus, both $\gamma$ and $\tau_C$ are essentially functions
of $\sigma$. According to all the above findings, an approximative
evaluation of Eq.~(\ref{alltcl}) to second order reads
\begin{equation}
\dot{F}_q(t) \approx \beta^2 \, \Gamma_{2,q}(t) \, F_q(t) \, ,
\quad \beta^2 \, \Gamma_{2,q}(t) \approx -W \, \tilde{\Gamma}_2(t)
\, . \label{diffusion3a}
\end{equation}
This implies for $t >\tau_C$
\begin{equation}
\dot{F}_q(t) \approx -W \, \gamma \, F_q(t) \, , \quad \tau_R
\equiv 1/(W \, \gamma) \, . \label{diffusion3}
\end{equation}
The comparison of Eq.~(\ref{diffusion3}) with (\ref{diffusion2})
clearly indicates diffusive behavior with a diffusion constant
$\kappa = \beta^2 \, \gamma$. Due to the independence of $\gamma$
from $n$, $N$ the pertinent diffusion constant for arbitrarily
large systems may be quantitatively inferred from the
diagonalization of a finite, e.g., ``$30 \times 30$''-layer.

\begin{figure}[htb]
\centering
\includegraphics[width=5.5cm]{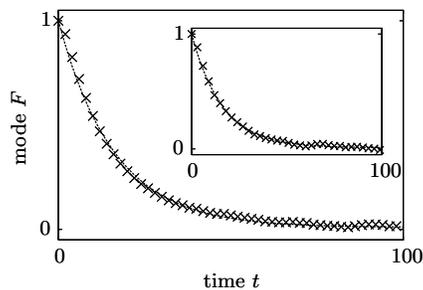}
\caption{Time evolution of modes $F_q$ with $q = 2 \pi / N$, the
longest wavelength. Parameters: $n = 30$, $\sigma = 1$, $N = 10$,
$\beta = 0.24$ (Inset: $N = 42$, $\beta = 1$). In both cases the
numerical result (crosses) is an exponential decay which clearly
indicates diffusive transport behavior and well agrees with the
TCL2 result (continuous curve), see text for details.}
\label{solution1}
\end{figure}

\begin{figure}[htb]
\centering
\includegraphics[width=5.5cm]{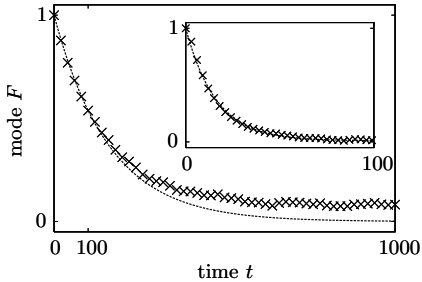}
\caption{Time evolution of a mode $F_q$ with $q = 2 \pi / N$, the
longest wavelength (Inset: $q = \pi$, the shortest wavelength).
Parameters: $n = 30$, $\sigma = 1$, $N = 10$, $\beta = 0.08$. The
TCL2 result (continuous curve) fails to describe the dynamics
correctly for sufficiently long wavelengths.} \label{solution2}
\end{figure}

To check this theory, we exemplarily present some results here.
For, e.g., $\sigma = 1$ and $n = 30$ we numerically find $\tau_C
\approx 10$ and $\gamma \approx 2.9$. Thus, additionally choosing
$\beta = 0.24$ and considering the longest wavelength mode in a $N
= 10$ system ($q = \pi / \, 5$), we find $W \gamma \approx 0.064$
[cf.~Eq.~(\ref{diffusion3})]. This corresponds to a ratio $\tau_R
/ \, \tau_C \approx 1.6$, that is, $\tau_R > \tau_C$, which
justifies the replacement of Eq.~(\ref{diffusion3a}) by
(\ref{diffusion3}) [see also the discussion of this issue in the
following paragraph]. And indeed, for the dynamics of $F_q(t)$ we
get an excellent agreement of the theoretical prediction based on
Eq.~(\ref{diffusion3}) with the numerical solution of the full
time-dependent Schr\"odinger equation (see Fig.~\ref{solution1}).
Note that this solution is obtained by the use of exact
diagonalization. Naturally interesting is the ``isotropic'' case
of $\beta= 1$. Keeping $\sigma = 1$, one has to go to the longest
wavelength in a $N = 42$ system in order to keep the $W$ from the
former example unchanged. If our theory applies, the decay curve
should be the same, which indeed turns out to hold (see inset of
Fig.~\ref{solution1}). Note that the integration in this case
already requires  approximative numerical integrators like, e.g.,
Suzuki-Trotter decompositions \cite{steinigeweg2006-1}. A
numerical integration of systems with larger $N$ rapidly becomes
unfeasible but an analysis based on Eq.~(\ref{diffusion3}) may
always be performed.

So far, we characterized the dynamics of the diffusive regime. We
turn towards an investigation of its size now. Obviously, the
replacement of Eq.~(\ref{diffusion3a}) by (\ref{diffusion3}) is
only self-consistent for $\tau_R > \tau_C$, i.e, if the relaxation
time is larger than the correlation time. This will possibly break
down for some large enough $q$ (small enough $l$), which then
indicates the transition to the ballistic regime. Since the
crossover is expected at $ \tau_R \approx \tau_C$, we may hence
estimate the maximum diffusive $q_\text{max}$ as
[cf.~Eq.~(\ref{diffusion3})]
\begin{equation}
\label{wmax} W_\text{max} = W(q_\text{max}) = 1 / (\tau_C \,
\gamma) \, .
\end{equation}
It turns that in the regime of $W \approx W_\text{max}$ a
description according to Eq.~(\ref{diffusion3a}) still holds.
However, in this case $\tilde{\Gamma}_2(t)$ is no longer
essentially constant but linearly increasing during the relaxation
period. This corresponds to a diffusion coefficient $\kappa$ which
increases linearly in time, which in turn is a strong hint for
ballistic transport (cf.~Ref.~\cite{steinigeweg2007-1}). Thus,
this transition may routinely be interpreted as the transition
towards ballistic dynamics which is expected on a length scale
below some mean free path.

\begin{figure}[htb]
\centering
\includegraphics[width=6.5cm]{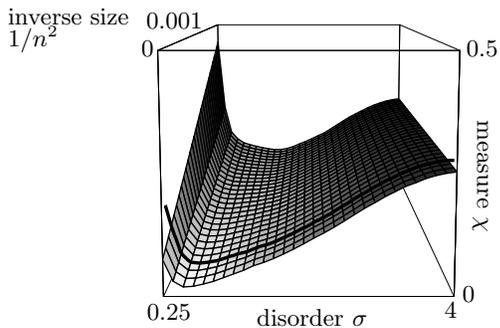}
\caption{Numerical results for the measure $\chi$ with respect to
the amount of disorder $\sigma$ and the inverse layer size $1 /
n^2$. An absolute minimum $\chi_{\text{min}} \approx 0.02$ is
found at $\sigma \approx 0.5$ in the limit of $n \rightarrow
\infty$. Note that only $10 \%$ of the whole area is extrapolated
(the area in front of the thick line).} \label{measure}
\end{figure}

In the following we intend to show that, in the limit of long
wavelengths, it is the influence of higher order terms in
Eq.~(\ref{alltcl}) that describes the deviation from diffusive
dynamics. To those ends we consider $L$, the ratio of second order
to fourth order terms
\begin{equation}
\label{ratio} L(t) \equiv \frac{\beta^4 \,
\Gamma_{4,q}(t)}{\beta^2 \, \Gamma_{2,q}(t)} \, .
\end{equation}
Whenever $L(t) \ll 1$, the decay is dominated by the second order
$\Gamma_{2,q}(t)$, which implies diffusive dynamics. It turns
unfortunately out that the direct numerical evaluation of
$\Gamma_{4,q}(t)$ is rather involved. However, a somewhat lengthy
calculation based on the techniques described in
Ref.~\cite{bartsch2007} shows that, for small $\Gamma_{4,q}(t)$,
the fourth order term assumes the same scaling in $\beta$, $q$ as
the second order term and may be approximated as
\begin{eqnarray}
\label{scale}
& \beta^4 \, \Gamma_{4,q}(t) \approx W^2 \, \tilde{\Gamma}_4(t) \, , \\
& \tilde{\Gamma}_4(t) \equiv t \, \Big[ 1/n^2 \sum \limits_i^{} \Big( \int
\limits_0^t \text{d} \tau \, \langle i | \, \hat{v}_0(t) \, \hat{v}_0(t') \, |
i \rangle \Big)^2 \!\!\! - \tilde{\Gamma}_2(t)^2 \Big] \nonumber \, ,
\end{eqnarray}
where $| i \rangle$ are eigenstates of $H_0$, i.e.,
$\tilde{\Gamma}_4(t)$ may be evaluated from considering some
``representative junction'' of only two layers, just as done for
$\tilde{\Gamma}_2(t)$. The calculation is based on the fact that
the interaction features Van Hove structure, that is, $V^2$
essentially is diagonal. We intend to give the details of this
calculation in a forthcoming publication. Here, we want to
concentrate on its results and consequences. [We should note that
all our data available from exact diagonalization is in accord
with a description based on Eqs.~(\ref{alltcl}),
(\ref{diffusion3a}) and especially (\ref{scale}). We should
furthermore note that $\tilde{\Gamma}_{4}(t)$, other than
$\tilde{\Gamma}_{2}(t)$, scales significantly with $n$, which
eventually gives rise to the $n$-dependence in
Fig.~\ref{measure}.] With Eq.~(\ref{scale}) we may rewrite
Eq.~(\ref{ratio}) as
\begin{equation}
L(W,t)= W^2 \, \frac{\tilde{\Gamma}_{4}(t)}{\tilde{\Gamma}_{2}(t)} \, .
\label{ratio1}
\end{equation}

This ratio turns out to be a monotonously increasing function in
$t$, which is not surprising, since lower order terms in
Eq.~(\ref{alltcl}) are expected to dominate at shorter times.
Thus, no visible deviation from the (diffusive) second order
description arises, as long as the decay is ``over'', before $
L(t)$ reaches some value on the order of a fraction of one. Since
the decay time scale is given by $\tau_R$, we are interested in
$L(W, \tau_R )$. If $L(\tau_R )$ is on the order of one, the
dynamics of the corresponding density wave in the corresponding
model must exhibit significant deviations from diffusive,
exponential decay. Because (apart from the model parameters $n$,
$\sigma$) $\tau_R$ only depends on $W$
[cf.~Eq.~(\ref{diffusion3})], we may now, exploiting
Eq.~(\ref{ratio1}), reformulate $L(W,\tau_R )$ only as a function
of $W$ and the model parameters $n$, $\sigma$ but without any
explicit dependence on $\beta$, $q$:
\begin{equation}
L[ \, W, \tau_R(W) \,] \equiv R(W) \label{depend}
\end{equation}
We call the above reformulation $R(W)$. It turns out that $R(W)$
decreases monotonously with $W$ such that the minimum $W$ for
which $R(W) < 1$ holds may be found from $R(W_\text{min})=1$. This
$W_\text{min}$ corresponds to the maximum wavelength beyond which
no diffusive behavior can be expected. Due to the fact that
$\tilde{\Gamma}_{4}(t)$, $\tilde{\Gamma}_{2}(t)$ and $\tau_R(W)$,
$\tau_C$ are numerically accessible, $W_\text{min}$,
$W_\text{max}$ can be computed for a wide range of model
parameters $n$, $\sigma$. In Fig.~\ref{measure} we display the
ratio $\chi \equiv W_\text{min} / W_\text{max}$ as a function of
those model parameters. With the approximation $W \approx \beta^2
\, q^2$ this ratio allows for the following interpretation:
\begin{equation}
\chi = \frac{W_\text{min}}{W_\text{max}} \approx
\frac{q^2_\text{min}}{q^2_\text{max}} =
\frac{l_\text{min}^2}{l_\text{max}^2} \label{interp}
\end{equation}
Hence $\sqrt{\chi}$ (which no longer depends on $\beta$) may be
viewed as the ratio of the shortest to the longest diffusive
wavelength, the smaller it is, the larger is the diffusive regime.

Obviously, for each layer size $n$ there is some disorder that
``optimizes'' the diffusive regime (minimizes $\chi$). But,
however, for $n = 30$ (back of Fig.~\ref{measure}) we find
$\sqrt{\chi_\text{min}} \approx 1/3$ at this optimum disorder,
which indicates about one diffusive wavelength. Exactly those
respective wavelengths have been chosen for the examples in
Fig.~\ref{solution1} and the inset in Fig.~\ref{solution2}, but
not in Fig.~\ref{solution2} itself. For all $\sigma$ and up to $n
= 100$ (which is about the limit for our simple numerics) $\chi$
clearly appears to be of the form $\chi(\sigma, n) = A(\sigma) /
n^2 + B(\sigma)$. Extrapolating this $1/n^2$-behavior yields a
suggestion for the infinite model $n = \infty$ (front of
Fig.~\ref{measure}). According to this suggestion, we find
$\sqrt{\chi_\text{min}} \approx 1/7$, again at optimum disorder.
This indicates a rather small regime of diffusive wavelengths,
even for the infinite system. We would like to repeat that these
findings apply at infinite temperature, i.e., the above small
diffusive regime is characterized by the fact that the dynamics
within it are diffusive at {\it all} energies.

\section*{Acknowledgments}
We sincerely thank H.-P.~Breuer and H.-J.~Schmidt for fruitful
discussions. Financial support by the Deutsche
Forschungsgemeinschaft is gratefully acknowledged.

\bibliographystyle{unsrt}


\end{document}